\documentclass[11pt]{article}
\usepackage[utf8]{inputenc}
\usepackage{amsfonts}
\usepackage{amsmath}
\usepackage{amssymb}
\usepackage{indentfirst}
\usepackage{graphicx}
\usepackage[colorlinks]{hyperref}
\usepackage{cite}

\usepackage{color}

\usepackage{microtype}
\usepackage[normalem]{ulem}

\numberwithin{equation}{section}
\allowdisplaybreaks

\def\ii{{\rm i}}

\definecolor{blue-violet}{rgb}{0.54, 0.17, 0.89}
\definecolor{PineGreen}{cmyk}{0.92, 0, 0.59, 0.25}
\definecolor{YellowOrange}{cmyk}{0, 0.42, 1, 0}

\usepackage{graphicx}

\usepackage{hyperref}

\usepackage{enumitem}

\begin{document}
\numberwithin{equation}{section}

\begin{center}
{\bf\LARGE On the role of torsion and higher forms in \\ off-shell supergravity} \\
\vskip 2 cm
{\bf \large Lucrezia Ravera$^{1,2}$}
\vskip 8mm
 \end{center}
\noindent {\small $^{1}$ \it DISAT, Politecnico di Torino, Corso Duca degli Abruzzi 24, 10129 Torino, Italy. \\
$^{2}$  \it INFN, Sezione di Torino, Via P. Giuria 1, 10125 Torino, Italy.
}

\begin{center}
\today
\end{center}

\vskip 2 cm
\begin{center}
{\small {\bf Abstract}}
\end{center}

We elaborate on the presence of a nonvanishing totally antisymmetric (super)torsion, equivalent to an axial vector, and higher forms in the ``new minimal'' and ``old minimal'' off-shell formulations of $\mathcal{N}=1$, $D=4$ supergravity. We adopt the geometric superspace approach and study both the geometric Lagrangian and the off-shell closure of the Bianchi identities in this framework, showing how the aforementioned axial vector torsion contributes to both the new and the old minimal set of auxiliary fields. In particular, to reproduce the old minimal set within the geometric setup, we also introduce two real auxiliary 3-form potentials.

\vfill
\noindent {\small{\it
    E-mail:  \\
{\tt lucrezia.ravera@polito.it}}}
   \eject

\tableofcontents

\noindent\hrulefill

\section{Introduction}\label{intro}

When we construct supersymmetric Lagrangians based on the fields of the on-shell representation multiplets, we observe that the algebra of supercharges closes only for field configurations satisfying the equations of motion. This is the reason why the associated actions are called on-shell.
There are cases where additional (auxiliary) fields can be appropriately added to the Lagrangian such that the supersymmetry algebra based on the full set of fields closes without implementing the equations of motion. In this case, the multiplet composed of the original and the auxiliary fields forms an off-shell representation of the supersymmetry algebra. The equations of motion of the auxiliary fields are algebraic, meaning that such fields are not dynamical and their degrees of freedom (d.o.f.) vanish on-shell.

Off-shell formulations of supersymmetric theories are believed to be necessary in view of quantization.
However, the off-shell matching of d.o.f. in supersymmetric theories is in general problematic for theories with eight supercharges or more, that is for extended supergravities in four dimensions and for  higher-dimensional models (dimensions larger than six).
On the other hand, $\mathcal{N}=1$ supergravity in four spacetime dimensions has two well-known minimal off-shell formulations, which are the so-called old minimal \cite{Wess:1978bu,Stelle:1978ye,Ferrara:1978em}\footnote{See also, e.g., \cite{Siegel:1988qu} for its relation with the superstring.} and new minimal \cite{Akulov:1976ck,Sohnius:1981tp,Gates:1981tu} ones. Both of them have 6 auxiliary bosonic d.o.f., in such a way that, taking into account the 6 off-shell d.o.f. of the vielbein $V^a_\mu$ (with $a,b,\ldots=0,1,2,3$, denoting anholonomic tangent space indices, and $\mu,\nu,\ldots=0,1,2,3$), the off-shell matching of bosonic and fermionic d.o.f. is realized (the gravitino $\psi^\alpha_\mu$ carries 12 d.o.f. off-shell).\footnote{In the following we will most frequently neglect the spinor index $\alpha=1,2,3,4$ to lighten the notation.}
More specifically, the auxiliary fields of old minimal supergravity are a complex scalar (2 d.o.f.) and a vector field not associated with a gauge symmetry (4 d.o.f.), while those of new minimal supergravity are an antisymmetric tensor $B_{\mu \nu}=-B_{\nu \mu}$ (3 d.o.f.) and a gauge vector field $A_\mu$ (3 d.o.f.). These two versions of minimal off-shell supergravity were later understood to be two different gauge fixings of $\mathcal{N}=1$, $D=4$ conformal supergravity \cite{Kaku:1978nz,Kaku:1978ea,Townsend:1979ki,Ferrara:1978rk,deWit:1981fh,Siegel:1978mj} (see also \cite{VanNieuwenhuizen:1981ae,Kugo:1982cu}).\footnote{Other gauge fixings are discussed in \cite{Gates:2000ae}.}

Several techniques have been developed which allow the construction of on-shell and, consequently, (whenever it is possible) off-shell actions given the on-shell one, most of which are based on superspace formulations. There are various approaches to superspace, based on different geometrical ideas, but they all have in common the fact that the notion of Grassmann variables (anticommuting c-numbers) as coordinates is essential. Superspace approaches are equivalent to ordinary spacetime ones, with the advantage that the superspace framework gives a better geometrical insight (see, for instance, \cite{VanNieuwenhuizen:1981ae,CDF1} for details on the geometry of superspace). In particular, on superspace one may have an understanding of supergravity analogous to that of general relativity on spacetime. All the approaches to supergravity in superspace involve a large symmetry group and a large number of fields, so that one eventually has to impose constraints in order to recover ordinary supergravity on spacetime. On the other hand, one can exploit the power of symmetry to construct
general theories in a systematic way.

In this scenario, the geometric approach to supergravity in superspace \cite{CDF2}, also known as the ``rheonomic approach'', gives a geometric interpretation to the supersymmetry transformations rules as diffeomorphisms in the fermionic directions of superspace. For recent comprehensive reviews of this topic we refer the reader to \cite{DAuria:2020guc,Castellani:2019pvh,Ravera:2018zvm}. 
The rheonomic approach is a powerful framework for the formulation of supergravity and rigid supersymmetric theories; It has proven to be a valuable asset in the construction of supersymmetric theories in various dimensions and amount of supersymmetry, providing a consistent formulation also in certain cases where a spacetime action description was not available.

In \cite{CDF2,DAuria:1982mkx} new minimal supergravity was developed in the geometric superspace approach, where the theory was shown to be based on a free differential algebra (an extension of the Maurer-Cartan equations to include higher-degree differential forms) involving, in particular, an auxiliary 2-form gauge potential $B^{(2)}$. In other words, the new minimal theory turned out to be the local theory of an appropriate free differential algebra. Moreover, in \cite{CDF3}, off-shell formulations of $\mathcal{N}=1$, $D=4$ supergravity have been revisited, always in the geometric approach, by studying the off-shell closure of the Bianchi identities (we will review this concept in Sec. \ref{revrheon}). Especially, starting from the $16 \oplus 16$ (16 bosonic d.o.f. $\oplus$ 16 fermionic d.o.f.) set of physical and auxiliary fields that one can use to describe matter coupled supergravity, a consistent truncation to the $12 \oplus 12$ sets of the new and old minimal models have been provided (the scenario is well-summarized in Table VI.9.VII of \cite{CDF3}). In particular, both the old and new minimal auxiliary sets involve a vector $t^a$. The latter, in the context of the old minimal model, carries 4 off-shell d.o.f., while in the new minimal case it is constrained to be divergenceless and therefore carries 3 off-shell d.o.f., as the antisymmetric tensor $B_{\mu \nu}$ of the original formulation. The fact that $t^a$ is associated with the auxiliary tensor $B_{\mu \nu}$ of the new minimal model emerges in the geometric approach as the super-field strength $F^{(3)}$ of the auxiliary 2-form $B^{(2)}$ appearing in the free differential algebra underlying the theory is parametrized (off-shell) by $t^a$. 
On the other hand, as also explained in \cite{CDF3}, in four spacetime dimensions the vector $t^a$ can be naturally identified with the dual of a (totally antisymmetric) torsion.\footnote{Here and in the following, with the term torsion we will mean, actually, supertorsion, just as with the term curvatures we will mean supercurvatures, that is super-field strengths.}

Driven by this construction, in the present paper we elaborate on the presence of such completely antisymmetric (that is, axial vector) torsion and higher forms in both the new minimal and the old minimal supergravity theories in the geometric superspace approach. In both cases, we study the off-shell closure of the Bianchi identities and the construction of the geometric off-shell action. In particular, we review the new minimal model in the geometric setup to highlight that, at least at the level of the Bianchi identities, the auxiliary 2-form is unnecessary if we endow the theory with a nonvanishing divergenceless axial vector torsion. Nevertheless, the auxiliary 2-form is useful to write the off-shell supergravity action (as done in \cite{CDF2,DAuria:1982mkx}). Consequently, we show that one can reproduce the old minimal theory within the geometric setup by considering as auxiliary fields an (unconstrained) axial vector torsion and two real 3-form potentials. This analysis allows to interpret the supertorsion, which in supergravity is naturally zero on-shell, as a useful auxiliary field to go off-shell, its relevance being particularly evident in the geometric formulation of the old minimal model.

The remainder of this paper is structured as follows: In Sec. \ref{revrheon} we summarize some important aspects of the rheonomic approach, in order to facilitate the understanding of the content of the subsequent sections. In Sec. \ref{nmsec} we review the geometric construction of the new minimal model, focusing on the role of torsion in this context and giving to its totally antisymmetric component a freshen interpretation. In Sec. \ref{omsec} we develop the old minimal supergravity model in the geometric approach to supergravity in superspace, showing that the old minimal set of auxiliary fields is reproduced considering a nonvanishing axial vector torsion and two real 3-forms as auxiliary fields. Section \ref{disc} is devoted to a final discussion and future developments. In Appendix \ref{appa} we collect our conventions and some useful formulas.

\section{Key aspects of the geometric superspace approach to supergravity}\label{revrheon}

Let us list in the following some key points regarding the rheonomic approach to supergravity (focusing on $\mathcal{N}=1$, $D=4$):

\begin{enumerate}[wide, labelindent=0pt]
\item The theory is given in terms of 1-form superfields $\mu^\mathcal{A}(x^\mu, \theta^\alpha)$ defined on superspace, $\mathcal{M}^{4|4}(x^\mu, \theta^\alpha)$, where $x^\mu$ are commuting bosonic coordinates and $\theta^\alpha$ are fermionic Grassmann coordinates.\footnote{The index $\mathcal{A}$ collectively labels all the 1-forms of the theory.} The set $\mu^\mathcal{A}$ includes the supervielbein $(V^a\,,\psi)$, which defines an orthonormal basis of superspace, with $V^a$ the bosonic vielbein and $\psi$ the gravitino 1-form, and the Lorentz spin connection $\omega^{ab}=-\omega^{ba}$. 

\item The set $\mu^{\mathcal{A}}$ defines the Maurer-Cartan 1-forms of the theory, which encode the algebraic structure of the given supergravity theory through their Maurer-Cartan equations. The latter provide, out of the vacuum, the definition of the supercurvature 2-forms $R^\mathcal{A}$, which are the field strengths of the theory (also referred to as super-field strengths).\footnote{In other words, the vacuum value of the supercurvatures ($R^\mathcal{A}=0$) gives the superalgebra in its dual Maurer-Cartan formulation.}

\item Supersymmetry transformations on spacetime are associated with diffeomorphisms in the fermionic ($\theta^\alpha$) directions of superspace. Supergravity theories are formulated from the condition of invariance under ``general super-coordinate transformations'', generalizing to superspace the geometric description of general relativity in terms of spacetime diffeomorphisms. 

\item The superfield 1-forms $\mu^\mathcal{A}$, together with their field strengths $R^\mathcal{A}$, are functions of all the coordinates of superspace, and they are related to the corresponding spacetime quantities by the restriction $\theta=d\theta=0$. In order for the theory on superspace to have the same physical content as the theory
on spacetime, some constraints (named ``rheonomic constraints'' in \cite{CDF2}) have to be imposed on the supercurvatures. \\
More precisely, the supercurvatures can be actually expressed in two different ways, that have
to be equivalent: They are defined from their symmetry properties and have to satisfy consistency constraints given by the closure of Bianchi identities; On the other and, being 2-forms in superspace, they can also be expanded along the supervielbein basis $(V^a \,, \psi)$ of superspace,
\begin{equation}
    R^\mathcal{A} = {R^\mathcal{A}}_{ab} V^a \wedge V^b + {R^\mathcal{A}}_{a \alpha} V^a \wedge \psi^\alpha + {R^\mathcal{A}}_{\alpha \beta} \psi^\alpha \wedge \psi^\beta \,,
\end{equation}
where the superspace tensors ${R^\mathcal{A}}_{ab}$ are referred to as inner components, while the ones which appear in the decomposition along at least one fermionic direction, namely ${R^\mathcal{A}}_{a \alpha}$ and ${R^\mathcal{A}}_{\alpha \beta}$, are the outer components. 
The above equation gives the so-called rheonomic parametrization of the supercurvatures. \\
The components in the parametrization of the supercurvatures can be determined by requiring the supercurvatures to satisfy the corresponding Bianchi identities (or better, in this context, ``Bianchi relations'') also when expressed in terms of their parametrizations. One generally finds that the outer components of the supercurvatures have to be expressed as linear tensor combinations of the inner ones (which are actually known in the literature as supercovariant field-strengths). These conditions (rheonomic constraints) guarantee that no additional d.o.f. is introduced in the theory in superspace compared to those already present on spacetime.

\item The same conclusion can be reached from the study of the Lagrangian, which is constructed geometrically \cite{CDF2},\footnote{For the building rules of a geometric supergravity Lagrangian we refer the reader to \cite{CDF2}.} by decomposing the related field equations with respect to independent sectors along supervielbein polynomials in superspace. 

\item There are cases (in particular, in the following we will consider that of $\mathcal{N}=1$, $D=4$ supergravity) in which one can also add auxiliary fields,\footnote{In general, the off-shell formulation of supersymmetric theories may require an infinite number of auxiliary fields. Nowadays, the generally recognized most natural geometric framework for more complicated cases is harmonic superspace \cite{Galperin:2001seg}.} 
which allow the matching of the number of bosonic and fermionic off-shell d.o.f. and the off-shell closure of the supersymmetry algebra. When this happens, the Bianchi identities close without applying/implying the equations of motion.

\item The rheonomic parametrization of the supercurvatures also provides the supersymmetry transformation laws leaving invariant the spacetime Lagrangian up to boundary terms. Indeed, as the supersymmetry transformation laws of the fields in superspace can be read as diffeomorphisms generated by tangent vectors $\varepsilon=\bar{\varepsilon}Q$ in the fermionic directions of superspace ($Q_\alpha$ being the supersymmetry generators and $\varepsilon^\alpha$ an infinitesimal spinor to be identified with the supersymmetry parameter), they can be expressed in terms of Lie derivatives. This means that we can write the supersymmetry transformation of a generic 1-form superfield $\mathbf{\Phi}(x,\theta)$ as
\begin{equation}
    \delta_\varepsilon \mathbf{\Phi} = \ell_\varepsilon \mathbf{\Phi} = \imath_\varepsilon \left( \nabla \mathbf{\Phi} \right) + \nabla \left( \imath_\varepsilon \mathbf{\Phi} \right) \,,
\end{equation}
where $\nabla$ generically denotes the covariant derivative with respect to the tensorial structure of
the given superfield $\mathbf{\Phi}$ and $\imath_\varepsilon$ is the contraction operator along odd directions of superspace with parameter $\varepsilon$ (in particular, $\imath_\varepsilon \psi = \varepsilon$ and $\imath_\varepsilon \mu^\mathcal{A}=0$ for $\mu^\mathcal{A} \neq \psi$). In performing the contraction, one has to use for $\nabla \mathbf{\Phi}$ its rheonomic parametrization as a 2-form in superspace. The same argument can be naturally generalized to 0-forms and to higher-degree forms.

\item The action is obtained by integrating the (bosonic) superspace Lagrangian 4-form $\mathcal{L}[\mu^\mathcal{A}]$ on a generic bosonic hypersurface $\mathcal{M}^4 \subset \mathcal{M}^{4|4}$ immersed in superspace,
\begin{equation}\label{genac}
    \mathcal{S} =  \int_{\mathcal{M}^4\subset \mathcal{M}^{4|4}} \mathcal{L}[\mu^\mathcal{A}] \,.
\end{equation}

\item The Lagrangian is written in a background-independent (geometric) way, independent of the choice of a metric, and it is therefore invariant under general coordinate transformations in superspace. For this to be possible, the Lagrangian 4-form has to be entirely expressed as wedge product of differential forms and their exterior derivatives. For this reason, in particular, the kinetic terms have to be written at first-order, thus avoiding the use of the Hodge dual of the field strengths. One can exploit general super-coordinate transformations to freely choose any $\mathcal{M}^4 \subset \mathcal{M}^{4|4}$ as the bosonic submanifold of integration in superspace, since any local deformation of the integration manifold can be reabsorbed by a superdiffeomorphism.

\item The invariance of the action does not coincide with the invariance (modulo total divergences) of the Lagrangian, and the diffeomorphisms in superspace are an off-shell invariance of the general geometrical action \eqref{genac} if \cite{CDF2}
\begin{equation}
    d \left( \mathcal{L}[\mu^\mathcal{A}] \right) = 0 \,,
\end{equation}
that is, if $\mathcal{L}[\mu^\mathcal{A}]$ is a closed form in superspace. The latter is therefore a requirement that must be fulfilled to produce an off-shell supergravity action within the geometric formulation.

\item Gauge potentials described by $p$-forms ($p > 1$) and associated to $p$-index antisymmetric tensors, which are typically contained in supergravities
in $D \geq 4$, are accommodated into the theory in the framework of supersymmetric free differential algebras (FDAs) \cite{CDF2,DAuria:1982mkx,sullivan,DAuria:1982uck,dffdas,Castellani:1992sv,Castellani:1995gz} (FDAs are sometimes also referred to as Cartan integrable systems or Chevalley-Eilenberg Lie algebras cohomology framework in supergravity), an extension of the
Maurer-Cartan equations to involve higher-degree differential forms. \\
Strictly speaking, one can write ($p$+1)-cochains (Chevalley cochains) $\Omega^{i|(p+1)}$ in some representation ${D^i}_j$ of a Lie group, which are ($p$+1)-forms, in terms of the 1-forms at disposal. If these cochains are closed ($d\Omega^{i|(p+1)}=0$), they are called cocycles. If a cochain is exact, it is called a coboundary. Of particular interest are those cocycles that are not coboundaries, which are elements of the Chevalley-Eilenberg cohomology.\footnote{If the closed cocycles are also coboundaries (exact cochains), then the cohomology class is trivial.} In the case in which this happens, we can introduce a $p$-form $A^{i|p}$ and write
the closed equation
\begin{equation}
    dA^{i|p} + \Omega^{i|(p+1)} = 0 \,,
\end{equation}
which, together with the Maurer-Cartan equations of the Lie algebra, constitutes the first germ of a free differential algebra, containing, besides the starting Maurer-Cartan 1-forms, also the new $p$-form $A^{i|p}$. This procedure can be now iterated taking as basis of new cochains the full set of forms and looking again for cocycles. The procedure can be iterated again and again, till no more cocycles can be found. In this way,
we obtain the largest free differential algebra associated with the initial Lie algebra. \\
In the supersymmetric case a set of nontrivial cocycles is generally present in superspace due to the existence of
Fierz identities obeyed by the wedge products of gravitino 1-forms. There, one further imposes the physical request that the FDA should be described in terms of fields in ordinary superspace, whose cotangent space is spanned by the supervielbein.
This corresponds to the physical request that the Lie superalgebra has a fiber bundle structure,
whose base space is spanned by the supervielbein, the rest of the fields spanning a fiber $H$. It follows that possible gauge fields and the Lorentz spin connection, belonging to $H$, must be excluded from the construction of the cochains. \\
Supersymmetric FDAs provide the algebraic (vacuum) structure underlying supergravity theories in $4 \leq D\leq 11$. To study the dynamics of the theory, one switches on the supercurvatures associated with the $p$-form gauge potentials appearing in the FDA, analogously to what is done for the 1-form fields.
\end{enumerate}

We are now ready to face the geometric superspace construction of minimal $D=4$ off-shell supergravity.

\section{New minimal supergravity with torsion in the rheonomic approach}\label{nmsec}

In the geometric superspace approach, the vacuum structure of new minimal supergravity is given by the following FDA \cite{CDF2,DAuria:1982mkx}:
\begin{equation}\label{FDAnmvacuum}
    \begin{split}
        & d \omega^{ab} - {\omega^a}_c \wedge \omega^{cb} = 0 \,, \\
        & \mathcal{D} V^a - \frac{\ii}{2} \bar{\psi} \wedge \gamma^a \psi = 0 \,, \\
        & \mathcal{D}\psi - \frac{\ii}{2} \gamma_5 \psi \wedge A = 0 \,, \\
        & dA = 0 \,, \\
        & d B^{(2)} - \frac{\ii}{2} \bar{\psi} \wedge \gamma_a \psi V^a = 0 \,,
    \end{split}
\end{equation}
where $\mathcal{D} \equiv d-\omega$ is the Lorentz-covariant derivative, $\omega^{ab}$ the Lorentz spin connection, $V^a$ the vielbein, $\psi$ the gravitino 1-form, $A$ a 1-form potential associated with a chiral charge, and $B^{(2)}$ a 2-form gauge potential. As we are going to see, the 2-form gauge potential $B^{(2)}$ will play the role of an auxiliary 2-form in superspace and will be particularly useful to write the off-shell action. The $d^2$-closure of \eqref{FDAnmvacuum} relies in the Fierz identity \eqref{fierz3psigammaa}.
Then, out of the vacuum one writes the supercurvatures associated with \eqref{FDAnmvacuum}, which are defined as
\begin{equation}\label{curvdef}
    \begin{split}
        \mathcal{R}^{ab} & \equiv d \omega^{ab} - {\omega^a}_c \wedge \omega^{cb} \,, \\
        T^a & \equiv \mathcal{D} V^a - \frac{\ii}{2} \bar{\psi} \wedge \gamma^a \psi \,, \\
        \rho & \equiv \mathcal{D}\psi - \frac{\ii}{2} \gamma_5 \psi \wedge A \,, \\
        F & \equiv dA \,, \\
        F^{(3)} & \equiv d B^{(2)} - \frac{\ii}{2} \bar{\psi} \wedge \gamma_a \psi V^a \,.
    \end{split}
\end{equation}
The corresponding Bianchi identities read\footnote{In the following, to lighten the notation, we will frequently omit writing the wedge product between differential forms.}
\begin{equation}
    \begin{split}
        & \mathcal{D} \mathcal{R}^{ab} = 0 \,, \\
        & \mathcal{D} T^a + \mathcal{R}^{ab} V_b - \ii \bar{\psi} \gamma^a \rho = 0 \,, \\
        & \mathcal{D} \rho + \frac{1}{4} \mathcal{R}^{ab} \gamma_{ab} \psi + \frac{\ii}{2} \gamma_5 \rho A - \frac{\ii}{2} \gamma_5 \psi F = 0 \,, \\
        & d F = 0 \,, \\
        & d F^{(3)} - \ii \bar{\psi} \gamma_a \rho V^a + \frac{\ii}{2} \bar{\psi} \gamma_a \psi T^a = 0 \,.
    \end{split}
\end{equation}
Taking into account this field content, in \cite{CDF2,DAuria:1982mkx} it was considered the action
\begin{equation}
    \mathcal{S}_{\text{nm}} =  \int_{\mathcal{M}^4\subset \mathcal{M}^{4|4}} \mathcal{L}_{\text{nm}} \,,
\end{equation}
where $\mathcal{L}_{\text{nm}}$ is the geometric 4-form superspace Lagrangian\footnote{As we shall see in the following, the parameter $\alpha$ along the first-order kinetic term will be constrained by the requirement $d\mathcal{L}_{\text{nm}}=0$. Here we also correct some misprints appearing in \cite{CDF2}.}
\begin{equation}
\begin{split}
    \mathcal{L}_{\text{nm}} & = \mathcal{R}^{ab} V^c V^d \epsilon_{abcd} + 4 \bar{\psi} \gamma_5 \gamma_a \rho V^a - 4 F \wedge B^{(2)} \\
    & + \alpha \left( t^a F^{(3)} V_a + \frac{1}{24} t_m t^m \epsilon_{abcd} V^a V^b V^c V^d \right) \,.
\end{split}
\end{equation}
The associated field equations are
\begin{align}
    \delta \omega^{ab} & : \quad 2 \epsilon_{abcd} T^c V^d = 0 \,, \\
    \delta t^a & : \quad \alpha F^{(3)} = \alpha F_{abc} V^a V^b V^c =  - \frac{\alpha}{3} \epsilon_{abcd} t^d V^a V^b V^c \,, \label{kinconstr} \\
    \delta V^a & : \quad 2 \mathcal{R}^{bc} V^d \epsilon_{abcd} - 4 \bar{\psi} \gamma_5 \gamma_a \rho - \frac{\ii \alpha}{2} t_b \bar{\psi} \gamma_a \psi V^b - \alpha t_a F^{(3)} + \frac{\alpha}{6} t_m t^m \epsilon_{abcd} V^b V^c V^d = 0 \,, \\
    \delta \bar{\psi} & : \quad 8 \gamma_5 \gamma_a \rho V^a - 4 \gamma_5 \gamma_a \psi T^a - \ii \alpha \gamma_a \psi V^a t_b V^b = 0 \,, \\
    \delta A & : \quad F^{(3)} = 0 \,, \\
    \delta B^{(2)} & : \quad - 4 F + \alpha V^a \mathcal{D} t_a - \alpha t_a \left( T^a + \frac{\ii}{2} \bar{\psi} \gamma^a \psi \right) = 0 \,.
\end{align}
Note that \eqref{kinconstr} is a kinematical equation. The above equations satisfy the vacuum condition, namely they admit the vacuum solution $\mathcal{R}^{ab}=T^a=\rho=F=F^{(3)}=t^a=0$, independently of the value of the parameter $\alpha$. Observe that the on-shell content of the theory is identical to the one of $\mathcal{N}=1$, $D=4$ pure supergravity. From the sector-by-sector study of the on-shell Bianchi identities (or, equivalently, of the equations of motion), one can prove that the on-shell rheonomic parametrization for the curvatures is
\begin{equation}
    \begin{split}
        \mathcal{R}^{ab} & = {\mathcal{R}^{ab}}_{cd} V^c V^d + \bar{\psi} {\Theta^{ab}}_c V^c \,, \\
        T^a & = 0 \,, \\
        \rho & = \rho_{ab} V^a V^b \,, \\
        F & = 0 \,, \\
        F^{(3)} & = 0 \,,
    \end{split}
\end{equation}
with
\begin{equation}
    \Theta_{ab|c} = - 2 \ii \gamma_{[a}\rho_{b]c} + \ii \gamma_{c}\rho_{ab} \,.
\end{equation}
The quantities ${\mathcal{R}^{ab}}_{cd}$ and $\rho_{ab}$ are the so-called supercovariant field strengths, and they differ, in general, from the spacetime projections of the supercurvatures. 
The tensor ${\mathcal{R}^{ab}}_{cd}$ and the tensor spinor $\rho_{ab}$ above satisfy the propagation equations
\begin{align}
    & {\mathcal{R}^{am}}_{bm} - \frac{1}{2} {\delta^a}_b {R^{mn}}_{mn} = 0 \,, \\
    & \gamma^m \rho_{mn} = 0 \,.
\end{align}

Now, regarding the off-shell theory, in \cite{CDF2} the off-shell parametrization of the supercurvatures was constructed by setting the supertorsion to zero. However, one can write a more general off-shell parametrization in which the supertorsion is nonvanishing, completely antisymmetric, and given in terms of an auxiliary axial vector field $t^a$ as\footnote{In \cite{CDF3} a similar argument was considered, introducing such supertorsion with a parameter $\kappa_2$, which, in particular, can be set to zero (here, instead, we fix $\kappa_2=1$). However, there the role of torsion was not considered as pivotal in the geometric off-shell construction there, since the field $t^a$ already appears in the parametrization of the other supercurvatures (including $F^{(3)}$). Here we take a slightly different point of view with respect to the one of \cite{CDF2,DAuria:1982mkx,CDF3}, considering the axial vector torsion as the actual fundamental auxiliary field to reproduce the new minimal set of auxiliary fields, and the 2-form $B^{(2)}$ just as an auxiliary 2-form useful to write rather straightforwardly the geometric off-shell Lagrangian (moreover, here we fix some typos appearing in \cite{CDF3}). The prominent role of torsion will be more evident in the old minimal case.}
\begin{equation}\label{totastor}
    T^a = {T^a}_{bc} V^b V^c = \epsilon^{abcd} t_d V_b V_c \,, \quad T_{abc} = T_{[abc]} = \epsilon_{abcd} t^d \,,
\end{equation}
with $[t_a]=L^{-1}$.
In this case, the complete off-shell parametrization is
\begin{equation}\label{offshellparanm}
    \begin{split}
        \mathcal{R}^{ab} & = {\mathcal{R}^{ab}}_{cd} V^c V^d + \bar{\psi} {\Pi^{ab}}_c V^c - \ii \bar{\psi} \gamma_d \psi \epsilon^{abcd}t_c \,, \\
        T^a & = \epsilon^{abcd} t_d V_b V_c \,, \\
        \rho & = \rho_{ab} V^a V^b + \ii a \gamma_5 \psi t_a V^a \,, \\
        F & = F_{ab} V^a V^b + \bar{\psi} \gamma_5 \phi_a V^a + \ii (1-a) \bar{\psi} \gamma_a \psi t^a \,, \\
        F^{(3)} & = - \frac{1}{3} \epsilon_{abcd} t^d V^a V^b V^c \,,
    \end{split}
\end{equation}
with
\begin{equation}\label{thetaphi}
    \begin{split}
        \Pi_{ab|c} 
        & = - 2 \ii \gamma_{[a} \sigma_{b]c} + \ii \gamma_c \sigma_{ab} + 2 \ii \gamma_{ab} \sigma_c + 2 \ii \gamma_{c[a} \sigma_{b]} - 2 \ii \delta_{c[a} \sigma_{b]} - 4 \ii \gamma_{abc} \sigma + 4 \ii \delta_{c[a} \gamma_{b]} \sigma \,, \\
        \phi_a & = 2 (1-a) \sigma_a - 6 (1+a) \gamma_a \sigma \,, 
    \end{split}
\end{equation}
where we have introduced the irreducible components $\sigma_{ab}$, $\sigma_a$, and $\sigma$ of the gravitino supercovariant field strength $\rho_{ab}$, that is
\begin{equation}\label{irrrho}
    \rho_{ab} = \sigma_{ab} - \gamma_{[a} \sigma_{b]} + \gamma_{ab} \sigma \,,
\end{equation}
with
\begin{equation}
    \gamma^a \sigma_{ab} = \gamma^a \sigma_a = 0 \,.
\end{equation}
Besides, let us stress that, having defined
\begin{equation}
    \mathcal{D}t_a = \mathcal{D}_b t_a V^b + \bar{\psi} \omega_a \,,
\end{equation}
we find
\begin{equation}
    \omega_a = - \frac{\ii}{2} \epsilon_{abcd} \gamma^b \rho^{cd} = \gamma_5 \sigma_a + 3 \gamma_5 \gamma_a \sigma 
\end{equation}
and the constraint\footnote{This can be easily checked, for instance, by looking at the $V^4$ sector of the Bianchi identity of $F^{(3)}$.}
\begin{equation}\label{datazero}
    \mathcal{D}^a t_a = 0 \,.
\end{equation}
Let us mention that the latter can be simply solved by
\begin{equation}
    t^a = \epsilon^{abcd} \partial_b B_{cd} \,,
\end{equation}
where $B_{ab}=-B_{ba}$ is a 2-index antisymmetric tensor (0-form). In particular, this implies $F^{(3)}=-2 \partial_a B_{bc} V^a V^b V^c$.

We stress that the whole mechanism continues to work well, and the same occurs for the counting of off-shell d.o.f., if we exclude the 3-form $F^{(3)}$ from the theory, keeping only the axial vector torsion $t^a$ (obeying \eqref{datazero}) and the 1-form gauge field $A$ as auxiliary fields. 

Finally, as we have recalled in Sec. \ref{revrheon}, in the geometric superspace approach the off-shell invariance of the action requires the geometric Lagrangian to be a closed 4-form.
In the case at hand, we have
\begin{equation}
    \begin{split}
        d \mathcal{L}_{\text{nm}} & = 2 \mathcal{R}^{ab} T^c V^d \epsilon_{abcd} + 4 \bar{\rho} \gamma_5 \gamma_a \rho V^a - 4 \bar{\psi} \gamma_5 \gamma_a \rho T^a - 4 F \wedge F^{(3)} \\
        & + \alpha \mathcal{D}t_a F^{(3)} V^a + \alpha t_a \left( \ii \bar{\psi} \gamma_b \rho V^b - \frac{\ii}{2} \bar{\psi} \gamma_b \psi T^b \right) V^a - \alpha t_a F^{(3)} \left( T^a +  \frac{\ii}{2} \bar{\psi} \gamma^a \psi \right) \\
        & + \frac{\alpha}{12} \mathcal{D} t_m t^m \epsilon_{abcd} V^a V^b V^c V^d + \frac{\alpha}{6} t_m t^m \epsilon_{abcd} \left( T^a + \frac{\ii}{2} \bar{\psi} \gamma^a \psi \right) V^b V^c V^d = 0 \,.
    \end{split}
\end{equation}
Once \eqref{offshellparanm} is used, this equation has two relevant projections, namely $\psi \psi VVV$ and $\psi VVVV$, both implying
\begin{equation}
    \alpha = -16 (1-a) \,.
\end{equation}
Therefore, the final form of the Lagrangian is
\begin{equation}\label{Lnmfin}
\begin{split}
    \mathcal{L}_{\text{nm}} & = \mathcal{R}^{ab} V^c V^d \epsilon_{abcd} + 4 \bar{\psi} \gamma_5 \gamma_a \rho V^a - 4 F \wedge B^{(2)} \\
    & -16 (1-a) \left( t^a F^{(3)} V_a + \frac{1}{24} t_m t^m \epsilon_{abcd} V^a V^b V^c V^d \right) \,.
\end{split}
\end{equation}
We are left with a parameter $a$ (appearing in \eqref{offshellparanm} and \eqref{Lnmfin}), which reflects the freedom we have of redefining the $\mathrm{U}(1)$ connection $A$,
\begin{equation}
    A' = A + 2 a t^a V_a \,.
\end{equation}
Thus, fixing the value of $a$ is equivalent to fixing a particular definition of $A$. Once the value of $a$ has been fixed by a particular choice of $A$, the parameter $\alpha$ is also fixed. If $a=1$, then $\alpha=0$, the off-shell rheonomic parametrizations boil down to
\begin{equation}
    \begin{split}
        \mathcal{R}^{ab} & = {\mathcal{R}^{ab}}_{cd} V^c V^d + \bar{\psi} {\Pi^{ab}}_c V^c - \ii \bar{\psi} \gamma_d \psi \epsilon^{abcd}t_c \,, \\
        T^a & = \epsilon^{abcd} t_d V_b V_c \,, \\
        \rho & = \rho_{ab} V^a V^b + \ii \gamma_5 \psi t_a V^a \,, \\
        F & = F_{ab} V^a V^b + \bar{\psi} \gamma_5 \phi_a V^a \,, \\
        F^{(3)} & = - \frac{1}{3} \epsilon_{abcd} t^d V^a V^b V^c \,,
    \end{split}
\end{equation}
where, in particular,
\begin{equation}\label{phia1}
    \phi_a=-12 \gamma_a \sigma \,,
\end{equation}
and the action takes the very simple and elegant form
\begin{equation}\label{anmfin}
        \mathcal{S}_{\text{nm}} = \int_{\mathcal{M}^4\subset \mathcal{M}^{4|4}} \Bigg ( \mathcal{R}^{ab} V^c V^d \epsilon_{abcd} + 4 \bar{\psi} \gamma_5 \gamma_a \rho V^a - 4 F \wedge B^{(2)} \Bigg ) \,.
\end{equation}
Thus, requiring the independence of the action from the specific choice of the spacetime section of superpace, namely off-shell supersymmetry invariance, fixes the action completely (which is done, in practice, by imposing the condition $d\mathcal{L}_{\text{nm}}=0$).

The $12 \oplus 12$ multiplet given by $\lbrace{ V^a \,, A \,, t^a \rbrace}\oplus \lbrace{ \psi \rbrace}$ satisfies an off-shell closed superalgebra.\footnote{One can also prove that the spinorial derivatives of ${R^{ab}}_{cd}$, $\rho_{ab}$, and $F_{ab}$ can be all expressed in terms of the fields we have already introduced.}
The supersymmetry transformations leaving the action constructed with \eqref{Lnmfin} invariant and closing the aforementioned off-shell algebra are
\begin{equation}
    \begin{split}
        \delta_\varepsilon \omega^{ab} & = \bar{\varepsilon} {\Pi^{ab}}_c V^c - 2 \ii \bar{\varepsilon} \gamma_d \psi \epsilon^{abcd} t_c \,, \\
        \delta_\varepsilon V^a & = \ii \bar{\varepsilon} \gamma^a \psi \,, \\
        \delta_\varepsilon \psi & = \mathcal{D} \varepsilon + \frac{\ii}{2} \gamma_5 \varepsilon A + \ii a \gamma_5 \varepsilon t_a V^a \,, \\
        \delta_\varepsilon A & = \bar{\varepsilon} \gamma_5 \phi_a V^a + 2 \ii (1-a) \bar{\varepsilon} \gamma_a \psi t^a \,, \\
        \delta_\varepsilon A^{(2)} & = \ii \bar{\varepsilon} \gamma_a \psi V^a \,,
    \end{split}
\end{equation}
where $\varepsilon$ is the supersymmetry parameter and ${\Pi^{ab}}_c$ and $\phi_a$ are given in \eqref{thetaphi}. Fixing $a=1$, one ends up with the following supersymmetry transformations leaving the action \eqref{anmfin} invariant:
\begin{equation}
    \begin{split}
        \delta_\varepsilon \omega^{ab} & = \bar{\varepsilon} {\Pi^{ab}}_c V^c - 2 \ii \bar{\varepsilon} \gamma_d \psi \epsilon^{abcd} t_c \,, \\
        \delta_\varepsilon V^a & = \ii \bar{\varepsilon} \gamma^a \psi \,, \\
        \delta_\varepsilon \psi & = \mathcal{D} \varepsilon + \frac{\ii}{2} \gamma_5 \varepsilon A + \ii \gamma_5 \varepsilon t_a V^a \,, \\
        \delta_\varepsilon A & = \bar{\varepsilon} \gamma_5 \phi_a V^a \,, \\
        \delta_\varepsilon A^{(2)} & = \ii \bar{\varepsilon} \gamma_a \psi V^a \,,
    \end{split}
\end{equation}
where $\phi_a$ is given by \eqref{phia1}.
Hence, we are left with the new minimal set of auxiliary fields $\lbrace{ A \,, t^a \rbrace}$, where $A=A_\mu dx^\mu$ is a gauge field and $t^a$ satisfies \eqref{datazero}.
Indeed, regarding the off-shell d.o.f. counting, we have $6+3+3=12$ off-shell bosonic d.o.f., given, respectively, by $V^a_\mu$, $A_\mu$, and $t^a$ (fulfilling $\mathcal{D}^a t_a = 0$), which match the 12 off-shell fermionic d.o.f. carried by the gravitino $\psi_\mu$.

It would be interesting to see whether it is possible to write geometrically the off-shell supergravity action without using $B^{(2)}$ and introducing, instead, $T^a$ explicitly into the action in such a way to keep $F$ present in the appropriate way. However, at first sight it seems this would require at least a term such as $A\wedge T_a \wedge V^a$, in which the gauge 1-form field $A$ appears in a ``bare'' form. We leave this point to a future investigation and move on, instead, to our proposal for the old minimal theory in the geometric superspace approach, where the key role of actual auxiliary field attributable to an axial vector torsion will be more evident.

\section{Geometric formulation of old minimal supergravity with torsion and 3-forms}\label{omsec}

In this section we develop an off-shell formulation of $\mathcal{N}=1$, $D=4$ supergravity in the rheonomic approach whose set of auxiliary fields coincide with the one of old minimal supergravity. In particular, we will reproduce the old minimal set of auxiliary fields considering a nonvanishing axial vector torsion and two real 3-forms as auxiliary fields.

Before moving on to our construction, let us mention that there already exist two variants of the old minimal formulation for $\mathcal{N} = 1$ supergravity
in four dimensions in which one or each of the two auxiliary scalars is replaced by the field strength of a gauge 3-form (these theories are known as 3-form
supergravity and complex 3-form supergravity, respectively \cite{Farakos:2017jme,Cribiori:2020wch,Kuzenko:2017vil}). However, at least to our knowledge, it was never shown that one can obtain old minimal supergravity just with  two real auxiliary 3-forms and torsion. The rheonomic approach makes all of this evident, highlighting the geometric character of the theory and the role of the auxiliary forms in this context.

Let us therefore consider the following vacuum FDA underlying the theory:
\begin{equation}\label{fdaom}
    \begin{split}
        & d \omega^{ab} - {\omega^a}_c \wedge \omega^{cb} = 0 \,, \\
        & \mathcal{D} V^a - \frac{\ii}{2} \bar{\psi} \gamma^a \psi = 0 \,, \\
        & \mathcal{D}\psi = 0 \,, \\
        & d A^{(3)}_- - \frac{1}{2} \bar{\psi} \gamma_{ab} \psi V^a V^b = 0 \,, \\
        & d A^{(3)}_+ - \frac{\ii}{2} \bar{\psi} \gamma_{ab} \gamma_5 \psi V^a V^b = 0 \,,
    \end{split}
\end{equation}
whose $d^2$-closure relies in the Fierz identities
\begin{align}
    & \bar{\psi} \gamma_{ab} \psi \bar{\psi} \gamma^a \psi = 0 \,, \\
    & \bar{\psi} \gamma_{ab} \gamma_5 \psi \bar{\psi} \gamma^a \psi = 0 \,.
\end{align}
The 3-forms $A^{(3)}_-$ and $A^{(3)}_+$ are real. One may observe that they can also be recast in a single complex 3-form
\begin{equation}
    A^{(3)} = A^{(3)}_+ + \ii A^{(3)}_- \,,
\end{equation}
such that, in vacuum,
\begin{equation}
    d A^{(3)} - \frac{\ii}{2} \bar{\psi} \gamma_{ab} \gamma_5 \psi V^a V^b - \frac{\ii}{2} \bar{\psi} \gamma_{ab} \psi V^a V^b = d A^{(3)} - \frac{1}{4} \bar{\psi} \gamma^{ab} \left( \mathbb{I} + \gamma_5 \right) \psi V^c V^d \epsilon_{abcd} = 0 \,.
\end{equation}
It is well-known that a 3-form does not give any dynamical contribution to a four-dimensional theory: In four spacetime dimensions, its derivative is a top form in spacetime, while its field strength (whose components along four vielbein must be proportional to the volume element in four dimensions) can be related to the presence of fluxes \cite{DAuria:2002qje,deWit:2003hq,Grana:2005jc,DAuria:2005muz,Fre:2006zxd,deWit:2008ta,Samtleben:2008pe,Bielleman:2015ina}. The fact that, in $D=4$, 3-forms are nondynamical just predisposes them to be well-suited auxiliary fields for the construction of an off-shell supergravity theory.
Then, out of the vacuum, one can switch on the complex 4-form field strength of $A^{(3)}$,
\begin{equation}
    F^{(4)} \equiv d A^{(3)} - \frac{\ii}{2} \bar{\psi} \gamma_{ab} \gamma_5 \psi V^a V^b - \frac{\ii}{2} \bar{\psi} \gamma_{ab} \psi V^a V^b = d A^{(3)} - \frac{1}{4} \bar{\psi} \gamma^{ab} \left( \mathbb{I} + \gamma_5 \right) \psi V^c V^d \epsilon_{abcd} \,.
\end{equation}
The latter can be split into two real 4-forms $F^{(4)}_+$ and $F^{(4)}_-$ as follows:
\begin{equation}
    F^{(4)} = F^{(4)}_+ + \ii F^{(4)}_- \,.
\end{equation}
Hence, let us define the supercurvatures associated with \eqref{fdaom} as
\begin{equation}\label{curvdefom}
    \begin{split}
        \mathcal{R}^{ab} & \equiv d \omega^{ab} - {\omega^a}_c \wedge \omega^{cb} \,, \\
        T^a & \equiv \mathcal{D} V^a - \frac{\ii}{2} \bar{\psi} \gamma^a \psi \,, \\
        \rho & \equiv \mathcal{D}\psi \,, \\
        F^{(4)}_- & \equiv d A^{(3)}_- - \frac{1}{2} \bar{\psi} \gamma_{ab} \psi V^a V^b \,, \\
        F^{(4)}_+ & \equiv d A^{(3)}_+ - \frac{\ii}{2} \bar{\psi} \gamma_{ab} \gamma_5 \psi V^a V^b \,.
    \end{split}
\end{equation}
The corresponding Bianchi identities are
\begin{equation}
    \begin{split}
        & \mathcal{D} \mathcal{R}^{ab} = 0 \,, \\
        & \mathcal{D} T^a + \mathcal{R}^{ab} V_b - \ii \bar{\psi} \gamma^a \rho = 0 \,, \\
        & \mathcal{D} \rho + \frac{1}{4} \mathcal{R}^{ab} \gamma_{ab} \psi = 0 \,, \\
        & d F^{(4)}_- - \bar{\psi} \gamma_{ab} \rho V^a V^b + \bar{\psi} \gamma_{ab} \psi T^a V^b = 0 \,, \\
        & d F^{(4)}_+ - \ii \bar{\psi} \gamma_{ab} \gamma_5 \rho V^a V^b + \ii \bar{\psi} \gamma_{ab} \gamma_5 \psi T^a V^b = 0 \,.
    \end{split}
\end{equation}
One can then prove that the rheonomic off-shell parametrizations of the supercurvatures \eqref{curvdefom} are
\begin{equation}
    \begin{split}
        \mathcal{R}^{ab} & = {\mathcal{R}^{ab}}_{cd} V^c V^d + \bar{\psi} \tilde{\Pi}^{ab}_{\phantom{ab}c} V^c - \ii \bar{\psi} \gamma_d \psi \epsilon^{abcd}t_c + 2 \ii \bar{\psi} \gamma_5 \gamma^{ab} \psi \phi^- - 2 \bar{\psi} \gamma^{ab} \psi \phi^+ \,, \\
        T^a & = \epsilon^{abcd} t_d V_b V_c \,, \\
        \rho & = \rho_{ab} V^a V^b - 2 \gamma_5 \gamma_a \phi^- \psi V^a + 2 \ii \gamma_a \phi^+ \psi V^a + \ii \gamma_5 \psi t_a V^a \,, \\
        F^{(4)}_- & = \phi^- \epsilon_{abcd} V^a V^b V^c V^d \,, \\
        F^{(4)}_+ & = \phi^+ \epsilon_{abcd} V^a V^b V^c V^d \,,
    \end{split}
\end{equation}
where we have introduced two auxiliary real scalar fields $\phi^-$ e $\phi^+$, equivalent to a complex scalar $\phi = \phi^+ + \ii \phi^-$ (with $[\phi]=L^{-1}$), which parametrize $F^{(4)}_-$ and $F^{(4)}_+$, respectively, and also appear in the outer components of $\mathcal{R}^{ab}$ and $\rho$. Note that also $t^a$, which parametrizes the off-shell supertorsion, appears in the outer components of the supercurvatures $\mathcal{R}^{ab}$ and $\rho$. Moreover, using the decomposition \eqref{irrrho}, we have
\begin{equation}\label{thetaom}
    \tilde{\Pi}_{ab|c} 
    = - 2 \ii \gamma_{[a} \sigma_{b]c} + \ii \gamma_c \sigma_{ab} + 2 \ii \gamma_{ab} \sigma_c + 2 \ii \gamma_{c[a} \sigma_{b]} - 2 \ii \delta_{c[a} \sigma_{b]} - 2 \ii \gamma_{abc} \sigma + 4 \ii \delta_{c[a} \gamma_{b]} \sigma \,.
\end{equation}
Furthermore, having defined
\begin{equation}
    \mathcal{D}t_a = \mathcal{D}_b t_a V^b + \bar{\psi} \tilde{\omega}_a \,,
\end{equation}
and
\begin{equation}
    \begin{split}
        d \phi^- & = \mathcal{D}_a \phi^- V^a + \bar{\psi} \lambda^- \,, \\
        d \phi^+ & = \mathcal{D}_a \phi^+ V^a + \bar{\psi} \lambda^+ \,,
    \end{split}
\end{equation}
we find
\begin{equation}
    \tilde{\omega}_a = \gamma_5 \sigma_a - \gamma_5 \gamma_a \sigma \,,
\end{equation}
together with
\begin{equation}
    \begin{split}
        \lambda^- & = - \frac{\ii}{12} \gamma_5 \gamma^{ab} \rho_{ab} = \ii \gamma_5 \sigma \,, \\
        \lambda^+ & = \frac{1}{12} \gamma^{ab} \rho_{ab} = - \sigma \,.
    \end{split}
\end{equation}
Let us highlight that, unlike what happens in the case of the new minimal construction, here the off-shell closure of the Bianchi identities does not imply any differential constraint on $t^a$.

We may then write the action
\begin{equation}
    \mathcal{S}_{\text{om}} =  \int_{\mathcal{M}^4\subset \mathcal{M}^{4|4}} \mathcal{L}_{\text{om}} \,,
\end{equation}
with\footnote{We consider the off-shell torsion, whose presence is intrinsic in the action, to be completely antisymmetric, namely of the form \eqref{totastor}.}
\begin{equation}
\begin{split}
    \mathcal{L}_{\text{om}} & = \mathcal{R}^{ab} V^c V^d \epsilon_{abcd} + 4 \bar{\psi} \gamma_5 \gamma_a \rho V^a + \beta_1 \left( \phi^- F^{(4)}_- - \frac{1}{2} {\phi^-}^2 \epsilon_{abcd} V^a V^b V^c V^d \right) \\
    & + \beta_2 \left( \phi^+ F^{(4)}_+ - \frac{1}{2} {\phi^+}^2 \epsilon_{abcd} V^a V^b V^c V^d \right) \,.
\end{split}
\end{equation}
The field equations of the theory are
\begin{align}
    \delta \omega^{ab} & : \quad 2 \epsilon_{abcd} T^c V^d = 0 \,, \\
    \delta V^a & : \quad 2 \mathcal{R}^{bc} V^d \epsilon_{abcd} - 4 \bar{\psi} \gamma_5 \gamma_a \rho - \beta_1 \phi^- \bar{\psi} \gamma_{ab} \psi V^b - 2 \beta_1 {\phi^-}^2 \epsilon_{abcd} V^b V^c V^d \nonumber \\
    & \phantom{:} \quad - \ii \beta_2 \phi^+ \bar{\psi} \gamma_{ab} \gamma_5 \psi V^b - 2 \beta_2 {\phi^+}^2 \epsilon_{abcd} V^b V^c V^d = 0 \,, \\
    \delta \bar{\psi} & : \quad 8 \gamma_5 \gamma_a \rho V^a - 4 \gamma_5 \gamma_a \psi T^a - \beta_1 \phi^- \gamma_{ab} \psi V^a V^b - \ii \beta_2 \phi^+ \gamma_{ab} \gamma_5 \psi V^a V^b = 0 \,, \\
    \delta A^{(3)}_- & : \quad \beta_1 d \phi^- = 0 \,, \\
    \delta A^{(3)}_+ & : \quad \beta_2 d \phi^+ = 0 \,, \\
    \delta \phi^- & : \quad \beta_1 F^{(4)}_- = \beta_1 \phi^- \epsilon_{abcd} V^a V^b V^c V^d \,, \label{kinf4m} \\
    \delta \phi^+ & : \quad \beta_2 F^{(4)}_+ = \beta_2 \phi^+ \epsilon_{abcd} V^a V^b V^c V^d \,. \label{kinf4p}
\end{align}
We observe that \eqref{kinf4m} and \eqref{kinf4p} are kinematical equations. The above equations satisfy the vacuum condition, namely they admit the vacuum solution $\mathcal{R}^{ab}=T^a=\rho=F^{(4)}_-=F^{(4)}_+=\phi^-=\phi^+=0$. 

The requirement of off-shell invariance of the action, that is
\begin{equation}\label{dLzeroom}
    \begin{split}
        d \mathcal{L}_{\text{om}} & = 2 \mathcal{R}^{ab} T^c V^d \epsilon_{abcd} + 4 \bar{\rho} \gamma_5 \gamma_a \rho V^a - 4 \bar{\psi} \gamma_5 \gamma_a \rho T^a \\
        & + \beta_1 d \phi^- F^{(4)}_- + \beta_1 \phi^- \left( \bar{\psi} \gamma_{ab} \rho V^a V^b - \bar{\psi} \gamma_{ab} \psi T^a V^b \right) \\
        & - \beta_1 \phi^- d\phi^- \epsilon_{abcd} V^a V^b V^c V^d - 2 \beta_1 {\phi^-}^2 \epsilon_{abcd} \left( T^a + \frac{\ii}{2} \bar{\psi} \gamma^a \psi \right) V^b V^c V^d \\
        & + \beta_2 d \phi^+ F^{(4)}_+ + \beta_2 \phi^+ \left( \ii \bar{\psi} \gamma_{ab} \gamma_5 \rho V^a V^b - \ii \bar{\psi} \gamma_{ab} \gamma_5 \psi T^a V^b \right) \\
        & - \beta_2 \phi^+ d\phi^+ \epsilon_{abcd} V^a V^b V^c V^d - 2 \beta_2 {\phi^+}^2 \epsilon_{abcd} \left( T^a + \frac{\ii}{2} \bar{\psi} \gamma^a \psi \right) V^b V^c V^d = 0 \,,
    \end{split}
\end{equation}
implies (as it can be proved by analyzing the two relevant projections $\psi \psi VVV$ and $\psi VVVV$ of \eqref{dLzeroom})
\begin{equation}
    \beta_1 = \beta_2 = - 16 \,.
\end{equation}
Note that all the parameters of the theory are thus fixed.

The $12 \oplus 12$ multiplet given by $\lbrace{ V^a \,, t^a \,, \phi^- \,, \phi^+ \rbrace}\oplus \lbrace{ \psi \rbrace}$ satisfies an off-shell closed superalgebra.
The supersymmetry transformations leaving invariant the final action
\begin{equation}\label{finalom}
    \begin{split}
        \mathcal{S}_{\text{om}} & = \int_{\mathcal{M}^4\subset \mathcal{M}^{4|4}} \Bigg [ \mathcal{R}^{ab} V^c V^d \epsilon_{abcd} + 4 \bar{\psi} \gamma_5 \gamma_a \rho V^a - 16 \Bigg ( \phi^- F^{(4)}_- - \frac{1}{2} {\phi^-}^2 \epsilon_{abcd} V^a V^b V^c V^d \\
        & + \phi^+ F^{(4)}_+ - \frac{1}{2} {\phi^+}^2 \epsilon_{abcd} V^a V^b V^c V^d \Bigg ) \Bigg ]
    \end{split}    
\end{equation}
and closing the off-shell algebra are
\begin{equation}
    \begin{split}
        \delta_\varepsilon \omega^{ab} & = \bar{\varepsilon} \tilde{\Pi}^{ab}_{\phantom{ab}c} V^c - 2 \ii \bar{\varepsilon} \gamma_d \psi \epsilon^{abcd} t_c + 4 \ii \bar{\varepsilon} \gamma_5 \gamma^{ab} \psi \phi^- - 4 \ii \bar{\varepsilon} \gamma^{ab} \psi \phi^+ \,, \\
        \delta_\varepsilon V^a & = \ii \bar{\varepsilon} \gamma^a \psi \,, \\
        \delta_\varepsilon \psi & = \mathcal{D} \varepsilon - 2 \gamma_5 \gamma_a \phi^- \varepsilon V^a + 2 \ii \gamma_a \phi^+ \varepsilon V^a + \ii \gamma_5 \varepsilon t_a V^a \,, \\
        \delta_\varepsilon A^{(3)}_- & = \bar{\varepsilon} \gamma_{ab} \psi V^a V^b \,, \\
        \delta_\varepsilon A^{(3)}_+ & = \ii \bar{\varepsilon} \gamma_{ab} \gamma_5 \psi V^a V^b \,,
    \end{split}
\end{equation}
where $\tilde{\Pi}^{ab}_{\phantom{ab}c}$ is given in \eqref{thetaom}.
We are thus left with the old minimal set of auxiliary fields $\lbrace{ t^a \,, \phi^- \,, \phi^+ \rbrace}$.
Indeed, regarding the off-shell d.o.f. counting, we have $6+4+1+1=12$ off-shell bosonic d.o.f., given, respectively, by $V^a_\mu$, $t^a$, $\phi^-$, and $\phi^+$, which match the 12 off-shell fermionic d.o.f. carried by the gravitino $\psi_\mu$.

Finally, we observe that the field equations of \eqref{finalom} can be rewritten as
\begin{align}
    \delta \omega^{ab} & : \quad T^a = 0 \,, \\
    \delta V^a & : \quad 2 \hat{\mathcal{R}}^{bc} V^d \epsilon_{abcd} - 4 \bar{\psi} \gamma_5 \gamma_a \hat{\rho} = 0 \,, \\
    \delta \bar{\psi} & : \quad 8 \gamma_5 \gamma_a \hat{\rho} V^a = 0 \,, \\
    \delta A^{(3)}_- & : \quad d \phi^- = 0 \,, \\
    \delta A^{(3)}_+ & : \quad d \phi^+ = 0 \,, \\
    \delta \phi^- & : \quad F^{(4)}_- = \phi^- \epsilon_{abcd} V^a V^b V^c V^d \,, \\
    \delta \phi^+ & : \quad F^{(4)}_+ = \phi^+ \epsilon_{abcd} V^a V^b V^c V^d \,,
\end{align}
where we have defined
\begin{equation}\label{newcurvdef}
    \begin{split}
        \hat{\mathcal{R}}^{ab} & \equiv \mathcal{R}^{ab} + 16 \left( {\phi^+}^2 + {\phi^-}^2 \right) V^a V^b - 2 \ii \bar{\psi} \gamma_5 \gamma^{ab} \psi \phi^- + 2 \bar{\psi} \gamma^{ab} \psi \phi^+ \,, \\
        \hat{\rho} & \equiv \rho + 2 \gamma_5 \gamma_a \phi^- \psi V^a - 2 \ii \gamma_a \phi^+ \psi V^a \,.
    \end{split}
\end{equation}
Let us mention that, regarding $\phi^-$ and $\phi^+$, one may write the solutions $\phi^-=\epsilon^{abcd}\partial_a A^-_{bcd}$ and $\phi^+=\epsilon^{abcd}\partial_a A^+_{bcd}$ (that is, $F^{(4)}_- = -24 \partial_a A^-_{bcd} V^a V^b V^c V^d$ and $F^{(4)}_+ = -24 \partial_a A^+_{bcd} V^a V^b V^c V^d$), where $A^-_{abc}$ and $A^+_{abc}$ are totally antisymmetric tensors.

On the other hand, in order to make the physical content of the on-shell theory clearer, it is first of all useful to decompose the four component spinor $\psi$ in eigenmodes $\psi_{\pm}$ of the matrix $\gamma_5$,
\begin{equation}
    \gamma_5 \psi = \pm \psi_{\pm} \,, \quad \psi = \psi_- + \psi_+ \,,
\end{equation}
where the projectors and the corresponding projections are given by
\begin{equation}
    \mathbb{P}_{\pm} \equiv \frac{1}{2} (\mathbb{I} \pm \gamma_5) \quad \Rightarrow \quad \mathbb{P}_{\pm} \psi = \psi_{\pm} \,, \quad \bar{\psi}_{\pm} = \bar{\psi} \mathbb{P}_{\pm} \,.
\end{equation}
Furthermore, in order to find chiral components of the fermionic expressions, we list the
following useful identities:
\begin{equation}
        \mathbb{P}_{\pm} \gamma_5 = \pm \mathbb{P}_{\pm} \,, \quad \mathbb{P}_{\pm} \gamma_a = \gamma_a \mathbb{P}_{\mp} \,, \quad \mathbb{P}_{\pm} \gamma_5 \gamma_a = \pm \gamma_a \mathbb{P}_{\mp} \,, \quad \mathbb{P}_{\pm} \gamma_{ab} = \gamma_{ab} \mathbb{P}_{\pm} \,.
\end{equation}
The supertorsion and the supercurvatures defined in \eqref{newcurvdef} can then be recast as follows:
\begin{equation}
    \begin{split}
         T^a & \equiv \mathcal{D} V^a - \ii \bar{\psi}_+ \gamma^a \psi_- \,, \\
         \hat{\mathcal{R}}^{ab} & \equiv \mathcal{R}^{ab} + 16 \phi \phi^* V^a V^b + 2 \phi^* \bar{\psi}_{+} \gamma^{ab} \psi_{+} + 2 \phi \bar{\psi}_{-} \gamma^{ab} \psi_{-} \,, \\
         \hat{\rho}_+ & \equiv \rho_+ - 2 \ii \phi \gamma_a \psi_- V^a \,, \\
         \hat{\rho}_- & \equiv \rho_- - 2 \ii \phi^* \gamma_a \psi_+ V^a \,, 
    \end{split}
\end{equation}
with
\begin{equation}
    \phi = \phi^+ + \ii \phi^- \,, \quad \phi^* = \phi^+ - \ii \phi^- \,, \quad \phi \phi^*=|\phi|^2 \,,
\end{equation}
and where we have also used the fact that
\begin{equation}
    \bar{\psi}_{\pm} \gamma^{ab} \psi_{\mp} = 0 \,, \quad \bar{\psi}_{\pm} \gamma^a \psi_{\pm} = 0 \,, \quad \bar{\psi}_+ \gamma^a \psi_- = \bar{\psi}_- \gamma^a \psi_+ \,.
\end{equation}
Now, since, on-shell, $\phi^-$ and $\phi^-$ reduce to constants, $\phi$ and $\phi^*$ can be treated as constant parameters (in particular, scale lengths). Therefore, one may perform the following rescaling:
\begin{equation}
    \omega^{ab} \rightarrow \omega^{ab} \,, \quad V^a \rightarrow \sqrt{\phi \phi^*} V^a \,, \quad \psi_+ \rightarrow \sqrt{\phi} \psi_+ \,, \quad \psi_- \rightarrow \sqrt{\phi^*} \psi_- \,.
\end{equation}
In this way, the curvatures above become
\begin{equation}
    \begin{split}
         T^a & \equiv \mathcal{D} V^a - \ii \bar{\psi}_+ \gamma^a \psi_- \,, \\
         \hat{\mathcal{R}}^{ab} & \equiv \mathcal{R}^{ab} + 16 \left( \phi \phi^* \right)^2 V^a V^b + 2 \phi \phi^* \bar{\psi}_{+} \gamma^{ab} \psi_{+} + 2 \phi \phi^* \bar{\psi}_{-} \gamma^{ab} \psi_{-} \,, \\
         \hat{\rho}_+ & \equiv \rho_+ - 2 \ii \phi \phi^* \gamma_a \psi_- V^a \,, \\
         \hat{\rho}_- & \equiv \rho_- - 2 \ii \phi \phi^* \gamma_a \psi_+ V^a \,, 
    \end{split}
\end{equation}
that is
\begin{equation}\label{adscurv}
    \begin{split}
         T^a & \equiv \mathcal{D} V^a - \frac{\ii}{2} \bar{\psi} \gamma^a \psi \,, \\
         \hat{\mathcal{R}}^{ab} & \equiv \mathcal{R}^{ab} + 4 e^2 V^a V^b + e \bar{\psi} \gamma^{ab} \psi \,, \\
         \hat{\rho} & \equiv \rho - \ii e \gamma_a \psi V^a \,,
    \end{split}
\end{equation}
where we have eventually restored the four component spinor $\psi$ and introduced the scale parameter
\begin{equation}
    e \equiv 2 \phi \phi^* = 2 |\phi|^2 > 0 \,.
\end{equation}
The supercurvatures \eqref{adscurv} are the $\rm{OSp}(1|4)$ ones. Hence, the on-shell content of the theory is equivalent to that of pure $\mathcal{N}=1$ supergravity with a negative cosmological constant $\Lambda=-12 e^2 = -3/\ell^2 = -48 |\phi|^4$, where $\ell$ is the AdS radius. To conclude, we report here the on-shell parametrization of the curvatures \eqref{adscurv} of AdS supergravity, which reads as follows:
\begin{equation}
    \begin{split}
        \hat{\mathcal{R}}^{ab} & = \hat{\mathcal{R}}^{ab}_{\phantom{ab}cd} V^c V^d + \bar{\psi} \hat{\Theta}^{ab}_{\phantom{ab}c} V^c \,, \\
        T^a & = 0 \,, \\
        \hat{\rho} & = \hat{\rho}_{ab} V^a V^b \,,
    \end{split}
\end{equation}
with
\begin{equation}
    \hat{\Theta}_{ab|c} = - 2 \ii \gamma_{[a}\hat{\rho}_{b]c} + \ii \gamma_{c}\hat{\rho}_{ab} \,,
\end{equation}
and looks formally the same as the one of ``flat'' (that is, without supersymmetric cosmological constant) $\mathcal{N}=1$, $D=4$ pure supergravity, but here the supercurvatures are the $\rm{OSp}(1|4)$ ones.

\section{Discussion}\label{disc}

From the analysis carried out in this paper it emerges that torsion and higher forms can play a prominent role in the construction of off-shell supergravity theories. In particular, regarding the formulation of the new minimal model within the geometric superspace setup, at the level of the Bianchi identities one can see that the auxiliary 2-form introduced in \cite{CDF2,DAuria:1982mkx} is unnecessary if we endow the theory with a nonvanishing divergenceless axial vector torsion. On the other hand, the auxiliary 2-form is particularly useful to write the off-shell action. Our interpretation of the respective role of the axial vector torsion and the auxiliary 2-form is slightly different with respect to the one given in \cite{CDF3}: There, the auxiliary 2-form was considered as the fundamental auxiliary field and the totally antisymmetric torsion was introduced only since, in any case, the field $t^a$ appears in the parametrization of the other curvatures; Here we take a somewhat different point of view, considering such axial vector torsion as the main auxiliary field and the auxiliary 2-form as a useful tool to properly write the off-shell action. On the other hand, the prominent role of a nonvanishing axial vector torsion as an auxiliary field, which in supergravity is naturally zero on-shell, is particularly evident in the geometric formulation we propose for the old minimal model. In this sense, one may interpret the torsion, in other circumstances maybe even with all its components, as a useful (set of) auxiliary field(s) to go off-shell: It can provide extra off-shell bosonic d.o.f. even if it can be reabsorbed in the spin connection, since, under this perspective, such d.o.f. can be interpreted as ``hidden'' in the latter. 
Moreover, always regarding the old minimal case, our geometric construction, which involves two real auxiliary 3-form potentials ($A^{(3)}_-$ and $A^{(3)}_+$, which can also be recast in a single complex 3-form), provides in a dynamical way the negative cosmological constant of the on-shell theory.
Let us mention that if, instead of considering a nonvanishing axial vector torsion, one tries by taking just the trace part of the supertorsion, which is a vector, to be nonvanishing, the result is that the latter cannot be used together with the auxiliary three forms $A^{(3)}_-$ and $A^{(3)}_+$ (that is to say, together with the auxiliary scalars $\phi^-$ and $\phi^+$ parametrizing the field strengths $F^{(4)}_-$ and $F^{(4)}_+$) to go off-shell, as it cannot give a suitable parametrization of $F^{(4)}_-$ and $F^{(4)}_+$. In any case, in the context of minimal $\mathcal{N}=1$, $D=4$ off-shell supergravity, the torsion components other than the totally antisymmetric one do not seem to play any role (in the sense, in particular, that they are not needed to match the bosonic and fermionic off-shell d.o.f.).

We argue that the new and old minimal supergravity theories with torsion could correspond to different gauge-fixed versions of four-dimensional $\mathcal{N}=1$ conformal supergravity with torsion. Some preliminaries on the formulation of the latter in the rheonomic approach have been discussed in \cite{DAuria:2021dth}, where, in particular, a gauge theory of the conformal group in four spacetime dimensions with a nonvanishing axial vector torsion was presented. At the purely bosonic level, the requirement of conformal invariance implies a differential condition (a Killing equation) on the axial vector torsion, and something similar is expected at the supersymmetric level. It is therefore of particular interest to probe the introduction of a nonvanishing torsion in the supersymmetric theory.

Furthermore, as both torsion and auxiliary higher forms can carry bosonic off-shell d.o.f., it appears that a complete study of the nontrivial cocycles of supergravity theories, including the disclosure of the hidden gauge structure underlying the associated FDAs following the lines of \cite{Andrianopoli:2016osu,Andrianopoli:2017itj} (see also \cite{Ravera:2021sly} and \cite{Penafiel:2017wfr,Ravera:2018vra}), may shed some light on the off-shell construction of more complicated (possibly, $\mathcal{N}$-extended) supergravity theories, maybe even for cases in which an off-shell formulation is not yet known. In particular, in the context of the hidden gauge structure underlying FDAs, some hints may come from the geometric formulation of supergravity based on the so called Maxwell superalgebra. Indeed, on one hand, the latter can be viewed as the hidden superalgebra underlying a supersymmetric FDA in four spacetime dimensions \cite{Ravera:2018vra} involving a 3-form gauge potential, and, on the other hand, its dual Maurer-Cartan formulation involves two extra 1-form fields (besides $V^a$ and $\psi$) which could be interpreted as auxiliary fields to go off-shell. They are a bosonic 1-form field $A^{ab}=-A^{ba}$ and a Majorana spinor 1-form $\xi$. Furthermore, following the approach of \cite{Andrianopoli:2014aqa} (see also \cite{Ipinza:2016con,Banaudi:2018zmh}), in \cite{Concha:2018ywv} it was proved that the inclusion of these fields by means of boundary terms in flat supergravity allows to restore supersymmetry when a nontrivial spacetime boundary is present. Subsequently, in \cite{Andrianopoli:2021rdk} their role as auxiliary fields for the bulk theory was elucidated: From the analysis of the equations of motion of the bulk plus boundary Lagrangian it emerged that, in this context, the field equations of these fields implement the Bianchi identities of Lorentz and supersymmetry, associated with $\omega^{ab}$ and $\psi$. The deepening of this study is left to future endeavours.

Although off-shell $\mathcal{N}=1$, $D=4$ supergravity is an already well-known theory, in this paper we have provided a new interpretation of torsion and higher forms in this simpler case. What we have mentioned above, and in particular the study of the hidden gauge structure of supersymmetric FDAs, could reveal useful in the off-shell formulation of more complicated theories, in the presence of hypermultiplets (and hence nonlocality), for which an infinite number of auxiliary fields is needed and where the most fruitful approach so far has been that of the harmonic superspace \cite{Galperin:2001seg}.

\section*{Acknowledgments}

The author wish to thank Laura Andrianopoli and Riccardo D'Auria for illuminating discussions and suggestions, and acknowledges the Department of Applied Science and Technology of the Polytechnic of Turin and the INFN for financial support.

\appendix

\section{Conventions and useful formulas}\label{appa}

Let us collect here our conventions and some useful formulas that we have used to derive the results obtained in the present paper.

We work with a mostly minus spacetime signature
$\eta_{ab}= \mathrm{diag}\left(+,-,-,-\right)$ and with Majorana spinors, satisfying $\bar \psi = \psi^T C$, where $C$ is the charge conjugation matrix. The symbol $\mathcal{D}\equiv d - \omega$ denotes the Lorentz-covariant derivative. In particular, we have
\begin{equation}
    \begin{split}
        \mathcal{R}^{ab} & \equiv d \omega^{ab} - {\omega^a}_c \wedge \omega^{cb} \,, \\
        \mathcal{D}V^a & \equiv dV^a-{\omega^a}_b \wedge V^b \,, \\
        \mathcal{D}\psi & \equiv  d \psi - \frac{1}{4} \omega^{ab} \wedge \gamma_{ab} \psi \,,
    \end{split}
\end{equation}
where $\mathcal{R}^{ab}$ is the Lorentz curvature and $\omega^{ab}=-\omega^{ba}$ the Lorentz spin connection.
The matrices $C \gamma_a$, $C\gamma_{ab}$, and $C \gamma_5 \gamma_{ab}$ are symmetric, while $C$, $C\gamma_5$, and $C\gamma_5 \gamma_{a}$ are antisymmetric. The gamma matrices in four spacetime dimensions obey
\begin{equation}
\begin{split}
& \lbrace{\gamma_a,\gamma_b \rbrace} = 2\eta_{ab}  \,, \quad  \left[\gamma_a,\gamma_b \right] = 2 \gamma_{ab} \,, \quad \gamma_5 \equiv - \ii \gamma_0 \gamma_1 \gamma_2 \gamma_3  \,, \\
& \gamma _0^{\dagger} = \gamma _0 \,, \quad \gamma _0 \gamma _i^{\dagger}  \gamma _0 = \gamma _i \quad (i=1,2,3) \,, \quad \gamma _5^{\dagger} = \gamma _5 \,, \\
& \epsilon _{abcd} \gamma^{cd} = 2 \ii \gamma_{ab} \gamma_5 \,, \quad \gamma_{ab} \gamma_5 = \gamma_5 \gamma_{ab} \,, \quad \gamma_a \gamma_5 = - \gamma_5 \gamma_{a} \,, \\
& \gamma_m \gamma^{ab}\gamma^m =0 \,, \quad \gamma_{ab}\gamma_m \gamma^{ab}=0   \,, \quad      \gamma_{ab}\gamma_{cd} \gamma^{ab}= 4 \gamma_{cd} \,, \quad \gamma_m \gamma^a \gamma^m = -2\gamma^a \,, \\
& \gamma^a \gamma_a = 4 \,, \quad \gamma_b \gamma^{ab} = - 3 \gamma^a \,, \quad \gamma^{ab} \gamma_b = 3 \gamma^a \,, \quad \gamma_{ab} \gamma^{ab} = - 12 \,, \\
& \gamma^{ab}\gamma^c = 2 \gamma^{[a}\delta^{b]c} + \gamma^{abc}  =2 \gamma^{[a}\delta^{b]c} + \ii \epsilon^{abcd}\gamma_5 \gamma_d \,, \\
& \gamma^c \gamma^{ab} = -2 \gamma^{[a}\delta^{b]c} + \gamma^{abc} = -2 \gamma^{[a}\delta^{b]c} + \ii \epsilon^{abcd}\gamma_5 \gamma_d \,, \\
& \gamma_{ab}\gamma_{cd} =  \ii \epsilon_{abcd}\gamma_5 -4{\delta^{[a}}_{[c}{\gamma^{b]}}_{d]} -2\delta^{ab}_{cd} \,.
\end{split}    
\end{equation}
We also report the following useful Fierz identities:
\begin{align}
& \psi \bar\psi =  \frac 14 \gamma_a\bar\psi \gamma^a\psi-\frac 18 \gamma_{ab}\bar\psi\gamma^{ab}\psi \,, \\
& \gamma_a\psi\bar\psi \gamma^a\psi = 0 \,, \label{fierz3psigammaa} \\
& \gamma_{ab}\psi\bar\psi\gamma^{ab}\psi = 0 \,,
\end{align}
together with the following irreducible representations:
\begin{equation}
\begin{split}
& \Xi^{a}_{(12)} \equiv \psi\bar\psi \gamma^a\psi \,, \\
& \Xi^{ab}_{(8)} \equiv \psi\bar\psi \gamma^{ab}\psi + \gamma^{[a} \Xi^{b]}_{(12)} \,,
\end{split}    
\end{equation}
which satisfy $\gamma_a \Xi^{a}_{(12)}=0$, $\gamma_a \Xi^{ab}_{(8)}=0$. Furthermore, we have
\begin{equation}
\gamma_{ab}\psi\bar\psi \gamma^a\psi =  - \gamma^a\psi\bar\psi \gamma_{ab}\psi = - \gamma_5\gamma^a \psi\bar\psi \gamma_{ab}\gamma_5\psi= \Xi^{(12)}_b \,.
\end{equation}
Finally, some useful spinor identities are
\begin{equation}
\begin{split}
& \bar{\psi} \xi = \left(-1\right)^{pq} \bar{\xi} \psi \,, \\
& \bar{\psi} (S) \xi = - \left(-1\right)^{pq} \bar{\xi} (S) \psi \,, \\
& \bar{\psi} (AS) \xi = \left(-1\right)^{pq} \bar{\xi} (AS) \psi \,, 
\end{split}    
\end{equation}
where $(S)$ is a symmetric matrix, $(AS)$ is an antisymmetric one, and $\psi$ and $\xi$ denote, respectively, a generic $p$-form spinor and a generic $q$-form spinor.

\end{document}